\documentclass[12pt]{article}
\input{epsf}
\newcommand{\define}{\buildrel \Delta \over =}
\newtheorem{theorem}{Theorem}
\newtheorem{lemma}{Lemma}

\begin{document}
\renewcommand{\textfraction}{0}

\title{Exact Hamming Distortion Analysis of  Viterbi
Encoded Trellis Coded Quantizers
\footnote{This research was
supported   by NSF Grants CCR-9902081 and CCR-9979443.}}
\author{\normalsize John Kieffer and Yu Liao\\
\small Dept. of Electrical \& Computer Engineering \\
[-5pt] \small University of Minnesota \\
[-5pt] \small Minneapolis, MN 55455\\
[-5pt] \small kieffer@ece.umn.edu}
\date{}
\maketitle
\thispagestyle{empty}
\begin{abstract}
Let $G$ be a finite strongly connected aperiodic  directed graph
in which each edge carries a label from a finite alphabet $A$.
Then $G$ induces a trellis coded quantizer for
encoding an alphabet $A$ memoryless source. 
 A source sequence of long finite length is encoded by finding a path in
 $G$ of that length whose sequence of labels is closest in Hamming
 distance
to the source sequence; finding the minimum distance path is a dynamic
programming problem that is solved using the Viterbi algorithm.
We show how a Markov chain can be used to obtain
a closed form expression for the asymptotic expected Hamming distortion
 per sample that results as the number
 of encoded source samples
increases without bound. \end{abstract}
\normalsize

\section{Introduction}
Let $G$ be a finite strongly connected aperiodic directed graph.
 We further suppose that $G$
is a labelled graph in the sense that each edge $e$ of $G$
carries a label $L(e)$ from a fixed finite alphabet $A$.\par

Let $d:A\times A \to \{0,1\}$ be the Hamming distance function
\[ d(a_1,a_2) = \left \{ \begin{array} {r@{\quad}l}
0, & a_1=a_2 \\
1, & a_1\not=a_2
\end{array} \right. \]
\noindent Let $\{X_i\}_{i=1}^{\infty}$ be an i.i.d. stochastic process with
alphabet $A$. In this paper, we develop a method to compute the number
\begin{equation}
D(G) \define \lim_{n\to\infty}n^{-1}E\left [ \min_{\pi_n} \{ \sum_{i=1}^n
d(X_i,L(e_i))\}\right ],
\label{eq1}
\end{equation}
where $\pi_n = (e_1,e_2,\cdots,e_n)$ varies over all paths of
length $n$ in $G$, consisting of edges $e_1,e_2,\cdots,e_n$.
The limit in (\ref{eq1}) exists by the subadditive ergodic theorem.\par

We discuss why it is important to know how to compute
$D(G)$. Let $R$ be a positive integer, and suppose
$G$ is chosen to have $2^R$ outgoing edges per vertex.
The labelled graph $G$ induces a trellis coded
quantizer \cite{marcellin} for encoding the samples
$X_1, X_2, \cdots, X_n$ for any positive integer $n$.
The induced trellis coded quantizer operates in two steps.
In the first step, the Viterbi algorithm is used to find an
optimal path
$(e_1,e_2,\cdots,e_n)$ in $G$
along which the total distortion $\sum_{i=1}^n d(X_i,L(e_i))$ is
minimized; the sequence $(L(e_1),L(e_2),\cdots,L(e_n))$
obtained by following the optimal path $(e_1,e_2,\cdots,e_n)$
is consequently the best quantized version of the
sequence $(X_1,X_2,\cdots,X_n)$.
In the second step, the trellis coded quantizer encodes the
optimal path into
$$B_n = nk+\lceil\log_2card({\cal V}_G)\rceil$$
bits, where
${\cal V}_G$ denotes the set of vertices of ${G}$
and $card({\cal V}_G)$ denotes the cardinality of the
set ${\cal V}_G$; these $B_n$ bits are transmitted
to the user. From the $B_n$ received bits,
the user determines
the optimal path $(e_1,e_2,\cdots,e_n)$ and then builds
the sequence $(L(e_1),L(e_2),\cdots,L(e_n))$ by following
this path; by construction, it is guaranteed that the sequence
$(L(e_1),L(e_2),\cdots,L(e_n))$ is the user's best possible estimate of the
sequence $(X_1,X_2,\cdots,X_n)$. The asymptotic compression rate
achieved by the trellis coded quantizer
as $n\to \infty$ is $R$ bits/sample, since
$$R = \lim_{n\to\infty} n^{-1}B_n.$$
Equation (\ref{eq1}) tells us that $D(G)$ is
the asymptotic expected distortion/sample generated by
the trellis coded quantizer as $n\to\infty$.
The information source with output $\{X_i\}_{i=1}^{\infty}$
has a distortion-rate function which may be computed via
the Blahut algorithm;  let $D(R)$ be the distortion-rate
function evaluated at rate $R$.
Then, $D({ G}) \geq D(R)$
and the trellis coded quantizer induced by ${ G}$
performs well if $D({ G}) \approx D(R)$.
It is an unsolved problem of source
coding theory to find in a fast manner a labelled graph ${ G}$ with
$2^R$ outgoing edges per vertex for which
$D({ G})$ is as close as desired to $D(R)$.
Since this paper allows one to explicitly compute $D({ G})$
for a fixed labelled graph ${ G}$, it is hoped that progress can
be made in the search for a ${ G}$ for which
$D({ G}) \approx D(R)$ by examining how
$D({ G})$ varies as one varies ${ G}$.\par

\section{The Method}

The Viterbi algorithm can be used to compute the minimum
over $\pi_n$ in (\ref{eq1});
we describe how this is done. For each vertex
$v$ of $ G$, let ${\cal I}(v)$ be the set of all
pairs $(v',e)$ such that $v'$ is a vertex of $ G$,
$e$ is an edge of ${ G}$, and $e$ leads from
$v'$ to $v$. Let ${\cal S}$
be the set of all
 vectors
$$s = (s(v): v\in {\cal V}_G)$$
with nonnegative  integer components. We define
a ``Viterbi transition operator''
$V : {\cal S}\times A \to {\cal S}$ as follows:
For each $s\in{\cal S}$ and $x\in A$,
 define
$V(s,x)$ to be the vector
$s_1 = (s_1(v): v\in {\cal V}_G)$
in ${\cal S}$ in which
$$s_1(v) = \min\{(v',e)\in {\cal I}(v) : s(v') + d(x,L(e))\}, \;\;\; v\in {\cal V}_G.$$
For each
$i=0,1,2,\cdots,n$, let $S^i$ be the random vector
$$S^i = (S^i(v) : v\in {\cal V}_G)$$
in which
$S^0$ is the zero vector and
$$S^i = V(S^{i-1},X_i), \;\;\; i=1,2,\cdots,n.$$
The Viterbi algorithm tells us that
\begin{equation}
\min S^n =  \min_{\pi_n} \{ \sum_{i=1}^n d(X_i,L(e_i))\},
\label{eq11}
\end{equation}
where $\min S^n$ denotes the minimum component of
vector $S^n$.
Notice that
$$ \min S^{i-1} \leq \min S^{i} \leq 1+\min S^{i-1},\;\;\; i=1,2,\cdots,n,$$
and therefore
\begin{equation}
\min S^n = {\mathrm card}(\{1\leq i \leq n : \min S^i\not = \min S^{i-1}\}).
\label{eqq}
\end{equation}
 For $i=0,1,\cdots,n$, let
$$\tilde{S}^i = (\tilde{S}^i(v) : v \in {\cal V}_G)$$
 be the vector in which
$$\tilde{S}^i(v) = S^i(v) - \min S^i, \;\;\; v\in {\cal V}_G.$$
Let $\tilde{V} : {\cal S}\times A \to {\cal S}$ be the ``reduced
Viterbi transition operator'' in which $\tilde{V}(s,x)$
is obtained from $V(s,x)$ by subtracting the minimum component of
$V(s,x)$ from each component of $V(s,x)$.
Then $\tilde{S}^0, \tilde{S}^1, \cdots, \tilde{S}^n$
are obtained from the recursion
\begin{eqnarray}
\tilde{S}^0 &  =&  zero\; vector\; in\; {\cal S}\nonumber\\
\tilde{S}^i& =& \tilde{V}(\tilde{S}^{i-1},X_i),
\;\;\;  i=1,2,\cdots,n\label{eq2}
\end{eqnarray}
Let $ {\cal S}_G$ be the set of all $s$ in $\cal S$
in which $s$ is the zero vector or in which
there exist $x_1,x_2,\cdots,x_n\in A$
for some positive integer $n$ such that
$s$ is obtained from the recursion
\begin{eqnarray*}
\tilde{s}^0 & =& zero\; vector\; in\; {\cal S}\\
\tilde{s}^i& =& \tilde{V}(\tilde{s}^{i-1},x_i), \;\;\;  i=1,2,\cdots,n\\
\tilde{s}^n& =&  s
\end{eqnarray*}
For each $i=1,2,\cdots,n$,
$$\min S^i\not = \min S^{i-1} \Leftrightarrow
 V(\tilde{S}^{i-1},X_i) \not \in {\cal S}_G.$$
This fact, coupled with (\ref{eqq}), allows us to conclude that
\begin{equation}
\min S^n = {\mathrm card}(\{1\leq i \leq n :
 V(\tilde{S}^{i-1},X_i) \not \in {\cal S}_G\}).
\label{eq10}
\end{equation}
\par
\begin{lemma} ${\cal S}_G$ is finite. Moreover, let $k$
be the smallest positive integer such that it is possible
to go from any vertex of ${ G}$ to any  vertex of ${ G}$
along a path
of length $k$. Then, if $s \in {\cal S}_G$, each component
of vector $s$  is $\leq k$.
\end{lemma}
\par
{\it Proof.} Let $s \in {\cal S}_G$. If
$s$ is the zero vector, we are done. Assume $s$ is not
the zero vector. We can therefore  find
$x_1,x_2,\cdots,x_n$ in $A$ for some positive integer $n$
such that if  $\tilde{s}^0$ is the zero vector in ${\cal S}$ and
$$\tilde{s}^i = \tilde{V}(\tilde{s}^{i-1},x_i), \;\;\; i=1,2,\cdots,n,$$
then $\tilde{s}^n=s$.
 Let ${s}^0,{s}^1,\cdots,{s}^n$ be the vectors
in which ${s}^0=\tilde{s}^0$ and
$${s}^i = V({s}^{i-1},x_i), \;\;\; i=1,2,\cdots,n.$$
We have
\begin{equation}
\tilde{s}^i  = {s}^i - \min{s}^i, \;\;\; i=1,2,\cdots,n.
\label{eq4}
\end{equation}
We show that the maximum and minimum components of ${s}^n$
differ by at most $k$, which by (\ref{eq4})
will complete the proof. (The maximum and minimum element
of $\tilde{s}^n=s$ will then differ by at most $k$; since
the minimum element of $s$ is $0$,
the maximum element will be at most $k$.)
 If $n\leq k$, this is obvious
because each component of ${s}^n$ is a sum of
$n$ Hamming distances. Assume $n>k$.
 Let $m$ be the minimum component of
${s}^n$ and let $M$ be the maximum component.
There is a path
$(e_1',e_2',\cdots,e_n')$ in ${ G}$ such that
$$\sum_{i=1}^n d(x_i,L(e_i')) = m.$$
Suppose $v$ is the vertex of $ G$
such that the maximum component  of ${s}^n$ is ${s}^n(v)$.
By changing the last $k$ edges of the path
$(e_1',e_2',\cdots,e_n')$,
we can find a path $(e_1'',e_2'',\cdots,e_n'')$ in $ G$
ending at $v$; this path  automatically
yields
$$\sum_{i=1}^n d(x_i,L(e_i'')) \leq m+k.$$
The number $M$ is the smallest sum
$$\sum_{i=1}^n d(x_i,L(e_i))$$
 along all paths $(e_1,e_2,\cdots,e_n)$ in $ G$ ending at $v$.
Therefore, $M\leq m+k$. This completes the proof of the lemma.\par
\vskip 0.25in

We are now able to describe our method for computing $D({ G})$.
For each $x\in A$, let $p(x)$ denote the
``source letter probability'' $\Pr[X_i=x]$.
 If $s_1, s_2$ belong to the state space ${\cal S}_G$,  a ``Markov
transition''  $s_1\to s_2$ is defined if and only if
$\tilde{V}(s_1,x)=s_2$ for some $x\in A$.
 \begin{theorem} There is a probability distribution
$(q(s) : s \in {\cal S}_G)$ such that
\begin{eqnarray}
q(s') &=& \sum_{\{(s,x)\in{\cal S}_G\times A : \tilde{V}(s,x)=s'\}} q(s)p(x),\;\;\; s'\in {\cal S}_G;
\label{eq80}\\
D({ G}) &=& \sum_{\{(s,x)\in{\cal S}_G\times A :  V(s,x)\not\in {\cal S}_G \}} q(s)p(x).
\label{eq6}
\end{eqnarray}
Furthermore, if the Markov transition relation $\to$ on ${\cal S}_G$
has only one closed class of irreducible states, then the
probability distribution $(q(s))$ satisfying (\ref{eq80}) is unique.
\end{theorem}
\par
{\it Proof.} From the fact that ${\cal S}_G$ is
finite and equations (\ref{eq1}) (\ref{eq11}) (\ref{eq2})
(\ref{eq10}), we can extract a limit to conclude that there
exists a ${\cal S}_G$-valued random variable
$\tilde{S}^0$ and an $A$-valued random variable $\tilde{X}_1$
such that
\begin{description}
\item[(i)] $\tilde{X}_1$ and $X_1$ have the same distribution
$(p(x) : x\in A)$.
\item[(ii)] $\tilde{S}^0$ and $\tilde{X}_1$ are independent.
\item[(iii)] Letting $\tilde{S}^1$ be the ${\cal S}_G$-valued
random variable defined by
$$\tilde{S}^1 \define \tilde{V}(\tilde{S}^0,\tilde{X}_1),$$
then $\tilde{S}^0$ and $\tilde{S}^1$ have the same distribution.
\item[(iv)] $D({ G}) = \Pr[  V(\tilde{S}^0,\tilde{X}_1)\not\in {\cal S}_G]$.
\end{description}
By property (ii),
\begin{equation}
\Pr[\tilde{S}^1=s'] = \sum_{s\in {\cal S}_G}
 \Pr[\tilde{S}^0 = s]\Pr[\tilde{V}(s,\tilde{X}_1)=s'].
\label{eq113}
\end{equation}
Let $(q(s) : s\in {\cal S}_G)$ be the common probability
distribution of  $\tilde{S}^0$ and $\tilde{S}^1$ (guaranteed
by property (iii)).
Equation (\ref{eq113}) becomes (\ref{eq80}) if one makes the
substitutions
\begin{eqnarray*}
\Pr[\tilde{S}^1=s'] &=& q(s'),\\
\Pr[\tilde{S}^0=s] &=& q(s),\\
\Pr[\tilde{V}(s,\tilde{X}_1)=s'] &=& \sum_{\{x\in A : \tilde{V}(s,x)=s'\}} p(x).
\end{eqnarray*}
(The third of these equations follows from property (i).)
From properties (ii),(iv),
$$D({ G}) = \Pr[V(\tilde{S}^0,\tilde{X}_1)\not\in {\cal S}_G] =
\sum_{s\in {\cal S}_G} \Pr[\tilde{S}^0=s]\Pr[  V(s,\tilde{X}_1)\not\in {\cal S}_G].$$
This equation reduces to (\ref{eq6}). The last part
of Theorem 1 concerning the uniqueness of $(q(s))$
is a well-known fact from Markov chain theory.

\section{Example: de Bruijn graph}
We take $G$ to be the labelled de Bruijn graph of size $8$
having $2$ incoming and $2$ outgoing edges per vertex,
 depicted in trellis form in the figure below; the
edge labels are chosen from the alphabet $A = \{a,b,c,d\}$  of size $4$.

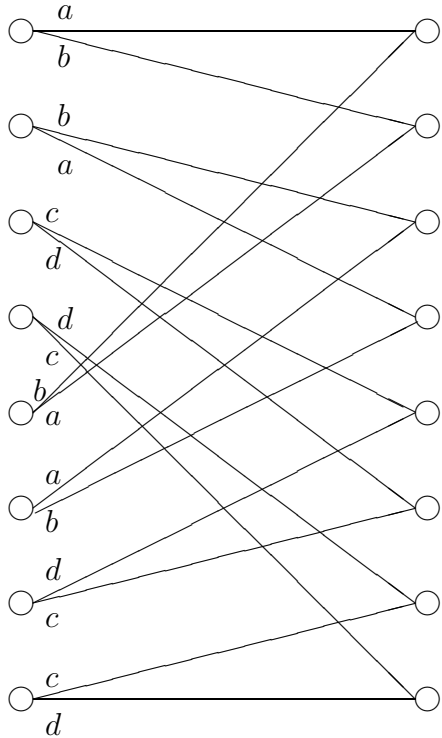
\begin{center}
\setlength{\unitlength}{3947sp}%
\begingroup\makeatletter\ifx\SetFigFont\undefined%
\gdef\SetFigFont#1#2#3#4#5{%
  \reset@font\fontsize{#1}{#2pt}%
  \fontfamily{#3}\fontseries{#4}\fontshape{#5}%
  \selectfont}%
\fi\endgroup%
\begin{picture}(2716,4590)(2543,-6136)
\thinlines
\put(2626,-1711){\circle{150}}
\put(2626,-2311){\circle{150}}
\put(2626,-2911){\circle{150}}
\put(2626,-3511){\circle{150}}
\put(2626,-4111){\circle{150}}
\put(2626,-4711){\circle{150}}
\put(2626,-5311){\circle{150}}
\put(2626,-5911){\circle{150}}
\put(5176,-1711){\circle{150}}
\put(5176,-2311){\circle{150}}
\put(5176,-2911){\circle{150}}
\put(5176,-3511){\circle{150}}
\put(5176,-4111){\circle{150}}
\put(5176,-4711){\circle{150}}
\put(5176,-5311){\circle{150}}
\put(5176,-5911){\circle{150}}
\put(2701,-1711){\line( 4,-1){2400}}
\put(2701,-2311){\line( 4,-1){2400}}
\put(2701,-2311){\line( 2,-1){2400}}
\put(2701,-2911){\line( 2,-1){2400}}
\put(2701,-2911){\line( 4,-3){2400}}
\put(2701,-3511){\line( 4,-3){2400}}
\put(2701,-3511){\line( 1,-1){2400}}
\put(2701,-5911){\line( 4, 1){2400}}
\put(2701,-5311){\line( 4, 1){2400}}
\put(2701,-5311){\line( 2, 1){2400}}
\put(2701,-4711){\line( 4, 3){2400}}
\put(2701,-4111){\line( 4, 3){2400}}
\put(2701,-4111){\line( 1, 1){2400}}
\put(2701,-5911){\line( 1, 0){2400}}
\put(2701,-1711){\line( 1, 0){2400}}
\put(2716,-4741){\line( 2, 1){2400}}
\put(2776,-3211){\makebox(0,0)[lb]{$d$}}
\put(2776,-4186){\makebox(0,0)[lb]{$a$}}
\put(2776,-4561){\makebox(0,0)[lb]{$a$}}
\put(2776,-5161){\makebox(0,0)[lb]{$d$}}
\put(2851,-1936){\makebox(0,0)[lb]{$b$}}
\put(2851,-1636){\makebox(0,0)[lb]{$a$}}
\put(2851,-2311){\makebox(0,0)[lb]{$b$}}
\put(2851,-2611){\makebox(0,0)[lb]{$a$}}
\put(2776,-2911){\makebox(0,0)[lb]{$c$}}
\put(2776,-3811){\makebox(0,0)[lb]{$c$}}
\put(2776,-4861){\makebox(0,0)[lb]{$b$}}
\put(2776,-5461){\makebox(0,0)[lb]{$c$}}
\put(2776,-5836){\makebox(0,0)[lb]{$c$}}
\put(2776,-6136){\makebox(0,0)[lb]{$d$}}
\put(2851,-3586){\makebox(0,0)[lb]{$d$}}
\put(2701,-4036){\makebox(0,0)[lb]{$b$}}
\end{picture}
\end{center}

We take our source output $\{X_i\}$ to consist of independent
equiprobable $A$-valued random variables. Let us
compute $D(G)$. \par

The state space ${\cal S}_G$ consists of $107$ vectors
of length $8$, which were found by computer search. It became
convenient to partition these $107$ vectors as follows:\newpage
\begin{eqnarray*}
\scriptstyle S_1&\scriptstyle =&\scriptstyle \{ 00000000\}\\
\scriptstyle S_2&\scriptstyle =&\scriptstyle \{00001111,\;11110000\}\\
\scriptstyle S_3&\scriptstyle =&\scriptstyle \{01101111,\;10011111,\;11110110,\;11111001\}\\
\scriptstyle S_4&\scriptstyle =&\scriptstyle \{11101122,\;11011122,\;22110111,\;22111011,\;01112211,\;10112211,\;11221110,\;11221101\}\\
\scriptstyle S_5&\scriptstyle =&\scriptstyle \{00001100,\;00000011,\;00110000,\;11000000\}\\
\scriptstyle S_6&\scriptstyle =&\scriptstyle \{22221210,\;22222101,\;22220121,\;22221012,\;10122222,\;01212222,\;21012222,\;12102222\}\\
\scriptstyle S_7&\scriptstyle =&\scriptstyle \{10111100,\;01111100,\;11010011,\;11100011,\;00111101,\;00111110,\;11001011,\;11000111\}\\
\scriptstyle S_8&\scriptstyle =&\scriptstyle \{22332101,\;22331210,\;33221012,\;33220121,\;10123322,\;01213322,\;21012233,\;12102233\}\\
\scriptstyle S_9&\scriptstyle =&\scriptstyle \{11101111,\;11011111,\;01111111,\;10111111,\;11110111,\;11111011,\;11111110,\;11111101\}\\
\scriptstyle S_{10}&\scriptstyle =&\scriptstyle \{22111001,\;22110110,\;11221001,\;11220110,\;10011122,\;01101122,\;10012211,\;01102211\}\\
\scriptstyle S_{11}&\scriptstyle =&\scriptstyle \{22221110,\;22221101,\;22220111,\;22221011,\;10112222,\;01112222,\;11012222,\;11102222\}\\
\scriptstyle S_{12}&\scriptstyle =& \scriptstyle \{21011122,\;12101122,\;10122211,\;01212211,\;22111012,\;22110121,\;11222101,\;11221210\}\\
\scriptstyle S_{13}&\scriptstyle =&\scriptstyle \{10011100,\;01101100,\;10010011,\;01100011,\;00111001,\;00110110,\;11001001,\;11000110\}\\
\scriptstyle S_{14}&\scriptstyle =&\scriptstyle \{10000011,\;01000011,\;00111000,\;00110100,\;11000001,\;00011100,\;00101100,\;11000010\}\\
\scriptstyle S_{15}&\scriptstyle =&\scriptstyle \{22110001,\;00012211,\;11220100,\;01001122,\;22110010,\;00102211,\;11221000,\;10001122\}\\
\scriptstyle S_{16}&\scriptstyle =&\scriptstyle \{10001111,\;01001111,\;11111000,\;11110100,\;11110010,\;11110001,\;00011111,\;00101111\}
\end{eqnarray*}
We call the sets $S_1, S_2, \cdots, S_{16}$ ``superstates''.
They are arrived at as follows. Let
$${\Pi} = \{\pi_1,\pi_2,\pi_3,\pi_4,\pi_5,\pi_6,\pi_7,\pi_8\}$$
 be the
group (under composition) of  permutations of ${\cal S}$
defined by
\begin{eqnarray*}
\pi_1(n_1,n_2,n_3,n_4,n_5,n_6,n_7,n_8) &=& (n_1,n_2,n_3,n_4,n_5,n_6,n_7,n_8)\\
\pi_2(n_1,n_2,n_3,n_4,n_5,n_6,n_7,n_8) &=& (n_2,n_1,n_4,n_3,n_6,n_5,n_8,n_7)\\
\pi_3(n_1,n_2,n_3,n_4,n_5,n_6,n_7,n_8) &=& (n_3,n_4,n_1,n_2,n_7,n_8,n_5,n_6)\\
\pi_4(n_1,n_2,n_3,n_4,n_5,n_6,n_7,n_8) &=&  (n_4,n_3,n_2,n_1,n_8,n_7,n_6,n_5)\\
\pi_5(n_1,n_2,n_3,n_4,n_5,n_6,n_7,n_8) &=& (n_5,n_6,n_7,n_8,n_1,n_2,n_3,n_4)\\
\pi_6(n_1,n_2,n_3,n_4,n_5,n_6,n_7,n_8) &=& (n_6,n_5,n_8,n_7,n_2,n_1,n_4,n_3)\\
\pi_7(n_1,n_2,n_3,n_4,n_5,n_6,n_7,n_8) &=&  (n_7,n_8,n_5,n_6,n_3,n_4,n_1,n_2)\\
\pi_8(n_1,n_2,n_3,n_4,n_5,n_6,n_7,n_8) &=& (n_8,n_7,n_6,n_5,n_4,n_3,n_2,n_1)
\end{eqnarray*}
If $s\in {\cal S}$, let $\Pi s$ denote the set
$$\Pi s \define \{\pi(s) : \pi \in \Pi\}.$$
The sets $\Pi s$, $s\in {\cal S}$, are called $\Pi$-{\it fibers}.
Any two  $\Pi$-fibers either coincide or have empty intersection. The
state space ${\cal S}_G$ is a union of $\Pi$-fibers; these
$\Pi$-fibers are the superstates $S_1,S_2,\cdots,S_{16}$.
We let ${\cal S}|\Pi$ denote the set of all $\Pi$-fibers.
We let ${\cal S}_G|\Pi$ denote the set of all $\Pi$-fibers
which are subsets of ${\cal S}_G$; the set
${\cal S}_G|\Pi$ is the same as the set of superstates
$\{S_1,S_2,\cdots,S_{16}\}$.
 \par
For any subgroup $\Pi'$ of $\Pi$, we can define the notion of
$\Pi'$-fiber in a similar fashion.
Let $\Pi'$ be the following subgroup of $\Pi$:
$$\Pi' \define \{\pi_1, \pi_2, \pi_5, \pi_6\}.$$
Let $S \in {\cal S}_G|\Pi$ be any superstate.
Let $S^1$ and $S^2$ be $\Pi'$-fibers whose union is $S$.
The following are true:
\begin{description}
\item[(a)] The set
\begin{equation}
\{V(s,x) : (s,x) \in S^1\times \{a,b\}\}\cup \{V(s,x) : (s,x) \in
S^2\times \{c,d\}\}
\label{eq201}
\end{equation}
is a $\Pi$-fiber.
\item[(b)] The set
\begin{equation}
\{V(s,x) : (s,x) \in S^2\times \{a,b\}\}\cup \{V(s,x) : (s,x) \in
S^1\times \{c,d\}\}
\label{eq202}
\end{equation}
is a $\Pi$-fiber.
\end{description}
Define $U(S,0)$ to be the $\Pi$-fiber (\ref{eq201}) and define
$U(S,1)$ to be the $\Pi$-fiber (\ref{eq202}). If $S$ is a
$\Pi$-fiber, let $S+1$ be the $\Pi$-fiber obtained by adding
$1$ to each component of each vector in $S$.
\par

From Theorem 1, we know that there is a probability
distribution $(q(s))$ on the state space ${\cal S}_{G}$
such that
\begin{eqnarray}
q(s') &=& \sum_{s \in {\cal S}_G} q(s)card(\{x\in A : V(s,x)\in \{s',s'+1\}\})/4, \;\; s'\in {\cal S}_G;
\label{eq90}\\
D(G) & =& \sum_{s \in {\cal S}_G} q(s)card(\{x\in A : V(s,x) \not \in {\cal S}_G\})/4.
\label{eq91}
\end{eqnarray}
Let $(q(S))$ be the probability distribution on
${\cal S}_{G}|\Pi$
such that
$$q(S) = \sum_{s \in S } q(s), \;\;\; S\in {\cal S}_G|\Pi  $$
Using properties (a)-(b) together with (\ref{eq90}) (\ref{eq91}),
one can show that
\begin{eqnarray}
q(S') &=& \sum_{S\in {\cal S}_G|\Pi} q(S)card(\{x\in \{0,1\} : U(S,x) \in \{S',S'+1\}\})/2,\;\; S'\in {\cal S}_G|\Pi;
\label{eq92}\\
D(G) & =& \sum_{S\in {\cal S}_G|\Pi} q(S)card(\{x\in \{0,1\} : U(S,x) \not \in {\cal S}_G|\Pi\})/2.
\label{eq93}
\end{eqnarray}
\par
We determined that:
$$\begin{array}{cccccc}
U(S_1,0) &=& S_2&\;\;\;\;\;\;\;\;\;\;\;\;\;\;\;\;U(S_1,1) &=& S_2\\
U(S_2,0) &=& S_3&\;\;\;\;\;\;\;\;\;\;\;\;\;\;\;\;U(S_2,1) &=& S_3\\
U(S_3,0) &=& S_4&\;\;\;\;\;\;\;\;\;\;\;\;\;\;\;\;U(S_3,1) &=& S_4\\
U(S_4,0) &=& S_6&\;\;\;\;\;\;\;\;\;\;\;\;\;\;\;\;U(S_4,1) &=& S_5+1\\
U(S_5,0) &=& S_2&\;\;\;\;\;\;\;\;\;\;\;\;\;\;\;\;U(S_5,1) &=& S_3\\
U(S_6,0) &=& S_8&\;\;\;\;\;\;\;\;\;\;\;\;\;\;\;\;U(S_6,1) &=& S_7+1\\
U(S_7,0) &=& S_9&\;\;\;\;\;\;\;\;\;\;\;\;\;\;\;\;U(S_7,1) &=& S_{10}\\
U(S_8,0) &=& S_8&\;\;\;\;\;\;\;\;\;\;\;\;\;\;\;\;U(S_8,1) &=& S_7+1\\
U(S_9,0) &=& S_{11}&\;\;\;\;\;\;\;\;\;\;\;\;\;\;\;\;U(S_9,1) &=& S_5+1\\
U(S_{10},0) &=& S_4&\;\;\;\;\;\;\;\;\;\;\;\;\;\;\;\;U(S_{10},1) &=& S_{12}\\
U(S_{11},0) &=& S_6&\;\;\;\;\;\;\;\;\;\;\;\;\;\;\;\;U(S_{11},1) &=& S_{13}+1\\
U(S_{12},0) &=& S_6&\;\;\;\;\;\;\;\;\;\;\;\;\;\;\;\;U(S_{12},1) &=& S_{14}+1\\
U(S_{13},0) &=& S_9&\;\;\;\;\;\;\;\;\;\;\;\;\;\;\;\;U(S_{13},1) &=& S_{15}\\
U(S_{14},0) &=& S_3&\;\;\;\;\;\;\;\;\;\;\;\;\;\;\;\;U(S_{14},1) &=& S_{16}\\
U(S_{15},0) &=& S_9&\;\;\;\;\;\;\;\;\;\;\;\;\;\;\;\;U(S_{15},1) &=& S_{10}\\
U(S_{16},0) &=& S_9&\;\;\;\;\;\;\;\;\;\;\;\;\;\;\;\;U(S_{16},1) &=& S_{10}
\end{array}$$
There is only one probability distribution satisfying
(\ref{eq92}). It is:
\begin{eqnarray*}
P(S_1)&=&0\\
P(S_2)&=&99/1809\\
P(S_3)&=&212/1809\\
P(S_4)&=&268/1809\\
P(S_5)&=&198/1809\\
P(S_6)&=&194/1809\\
P(S_7)&=&194/1809\\
P(S_8)&=&194/1809\\
P(S_9)&=&128/1809\\
P(S_{10})&=&112/1809\\
P(S_{11})&=&64/1809\\
P(S_{12})&=&56/1809\\
P(S_{13})&=&32/1809\\
P(S_{14})&=&28/1809\\
P(S_{15})&=&16/1809\\
P(S_{16})&=&14/1809
\end{eqnarray*}
From equation (\ref{eq93}),
\begin{eqnarray*}
D(G)& =& q(S_4)/2 + q(S_6)/2 + q(S_8)/2 + q(S_9)/2 + q(S_{11})/2
 + q(S_{12})/2\\
&=& 452/1809\\
& = & 0.2499\cdots
\end{eqnarray*}

\end{document}